\documentclass[twocolumn,showpacs,floats,floatfix,superscriptaddress,aps,pra]{revtex4-1}
\usepackage{amsfonts}
\usepackage{amssymb}
\usepackage{amsmath}
\usepackage{calc}
\usepackage{graphicx}
\usepackage{bm}

\usepackage[normalem]{ulem}
\usepackage{amsmath,amssymb}
\usepackage{multirow}
\usepackage{xcolor,soul}
\usepackage{srcltx}
\usepackage{hyperref,graphicx}

\def\be{ \begin{equation}}
\def\ee{ \end{equation}}
\def\bea{ \begin{eqnarray}}
\def\eea{ \end{eqnarray}}
\def\bse{ \begin{subequations}}
\def\ese{ \end{subequations}}
\def\bc{ \begin{center}}
\def\ec{ \end{center}}

\begin{document}

\author{Stefano Longhi$^{*}$} 
\affiliation{Dipartimento di Fisica, Politecnico di Milano and Istituto di Fotonica e Nanotecnologie del Consiglio Nazionale delle Ricerche, Piazza L. da Vinci 32, I-20133 Milano, Italy}
\email{stefano.longhi@polimi.it}

\title{Bloch oscillations in non-Hermitian lattices with trajectories in complex plane}
  \normalsize


%
\bigskip
\begin{abstract}
\noindent  

Bloch oscillations (BOs), i.e. the oscillatory motion of a quantum particle in a periodic potential, are one of the most striking effects of coherent quantum transport in the matter. In the semiclassical picture, it is well known that BOs can be explained owing to the periodic band structure of the crystal and the so-called 'acceleration' theorem: since in the momentum space the particle wave packet drifts with a constant speed without being distorted, in real space the probability distribution of the particle undergoes a periodic motion following a trajectory which exactly reproduces the shape of the lattice band. In non-Hermitian lattices with a complex (i.e. not real) energy band, extension of the semiclassical model is not intuitive. Here we show that the acceleration theorem holds for non-Hermitian lattices with a complex energy band only {\it on average}, and that the periodic wave packet motion of the particle in the real space is described by a trajectory in {\it complex} plane, i.e. it generally corresponds to reshaping and breathing of the wave packet in addition to a transverse oscillatory motion. The concept of BOs involving complex trajectories is exemplified by considering two examples of non-Hermitian lattices with a complex band dispersion relation, including the Hatano-Nelson tight-binding Hamiltonian describing the hopping motion of a quantum particle on a linear lattice with an imaginary vector potential and a tight-binding lattice with imaginary hopping rates.

\end{abstract}

\pacs{03.65.-w,  73.23.Ad, 03.65.Nk, 42.82.Et}

\maketitle

\section{Introduction}

Bloch oscillations (BOs), i.e. the oscillatory motion of a quantum particle in a periodic potential driven by a
constant force, are one of the most striking effects of coherent quantum transport in the matter. Originally studied in
the context of electrons in crystals \cite{r1,r2,r3,r3bis}, they have been observed  in a wide variety of different physical systems such as 
semiconductor superlattices \cite{r4,r5}, ultracold atomic gases \cite{r6,r7,r8}, optical and acoustical waves \cite{r9,r10,r11,r12,r13}. 
In the simplest semiclassical picture, BOs can be explained from the 
Ehrenfest equations of motion taking into account the periodic band structure of the crystal \cite{r14,r15,r16}. In particular, 
owing to the so-called 'acceleration theorem'  the momentum distribution of the particle moves in momentum space
at a constant speed while preserving its shape \cite{r2,r16,r17,r18}. As a result, for a wave packet with narrow momentum distribution
 the oscillation trajectory of the particle wave packet in physical space exactly reproduces the shape of the lattice band dispersion curve.
In the spectral domain, BOs are ascribed to the transition of
the energy spectrum from a continuous band with delocalized
Bloch eigenstates in absence of the external force to a
discrete ladder spectrum with localized Wannier-Stark (WS)
eigenstates \cite{r3bis} when the external force is applied and Zener tunneling
among different lattice bands is negligible \cite{r16,r18,r19}. Several studies have extended BOs to 
include the role of lattice disorder \cite{r20,r21}, nonlinearities \cite{r22}, lattice defects
and quasicrystals \cite{r23,r24}, inhomogeneous fields \cite{r25}, particle interaction and correlation \cite{r26,r27,r28,r29,r30}, BOs of nonclassical states of light \cite{r31,r32}, long-range hoppings \cite{r32bis}, etc.  \par
More recently, the onset of BOs have been investigated in lattices with a complex periodic potential \cite{r33,r34}, so-called complex crystals \cite{r35,r36} including those with parity-time ($\mathcal{PT}$) symmetry \cite{r37,r38}. The observation of BOs in complex crystals, reported in the landmark experiment of Ref.\cite{r39} using light waves in $\mathcal{PT}$-symmetric synthetic lattices, has raised the interest into the coherent transport properties in non-Hermitian lattices.
Some peculiar properties of non-Hermitian BOs  have been disclosed, such as the occurrence of complex WS ladders \cite{r33}, unidirectional BOs at the symmetry breaking transition \cite{r33,r39} arising from one-way Bragg scattering \cite{r36,r40}, and periodic BOs at exceptional points in spite of lattice truncation \cite{r41}.  However, it is not clear whether the 'acceleration theorem' and semiclassical picture of BOs for wave packets in ordinary lattices can be extended (and to what extent) to complex crystals. For the case where the imaginary part of the potential can be regarded as a perturbation to the real potential or whenever the band spectrum remains real in spite of the complex potential, the wave packet in real space undergoes an oscillatory trajectory like in the Hermitian case, except that BOs  are damped or amplified depending on the sign of the force \cite{r33}.  However, in strongly non-Hermitian potentials the band structure can become imaginary, and in this case it is not clear whether and how
 the semiclassical picture of BOs can be extended to account for a complex energy lattice band. Several examples of tight-binding lattice models that show a non-vanishing imaginary part of the band dispersion curve have been discussed in many works, including the Hatano-Nelson model describing the hopping motion of a quantum particle in a linear tight-binding lattice with an imaginary vector potential \cite{r33_1,r33_2,r33_3,r33_4}, the non-Hermitian extension of the Su-Schrieffer-Heeger tight-binding model \cite{r33_5,r33_6,r33_6bis}, $\mathcal{PT}$-symmetric binary superlattices \cite{r33,r33_4bis}, and the $\mathcal{PT}$-symmetric Aubry-Andre model \cite{r33_7}. Lattice bands with a non vanishing imaginary part could be realized in synthetic temporal optical crystals,  waveguide lattices, coupled-resonator optical waveguides, microwave resonator chains, etc. as discussed in several recent works \cite{r39,r41,r33_4,r33_4bis,r33_8,r33_9,r33_10}.\\  In this work we aim to investigate the onset of BOs in non-Hermitian lattices  with a non vanishing imaginary part of the band dispersion curve. Our analysis discloses that the ordinary wisdom of BOs  that uses a semiclassical description can not be trivially extended to non-Hermitian lattices when the band dispersion curve has a non vanishing imaginary part. In particular, we show that the acceleration theorem (i.e. the uniform drift in the momentum space of the wave packet distribution) holds only on average, whereas in real space  a particle wave packet with narrow spectral distribution undergoes a periodic motion {\it but}  following a closed orbit  in {\it complex} plane.  As a result, strong wave packet reshaping is generally observed within each BO cycle, in addition to the transverse oscillatory motion.  We exemplify the concept of BOs involving complex trajectories by considering the onset of BOs in two nearest-neighbor tight-binding lattice models with a complex energy band dispersion curve, including the Hatano-Nelson model describing the hopping motion of a quantum particle on a linear lattice with an imaginary vector potential. In such a case, it is shown that the BO motion of a particle wave packet corresponds to an elliptical orbit in the complex plane and the trajectory followed by a wave packet in real space can not be merely predicted by the real part solely (group velocity) of the energy dispersion relation.\\
 The paper is organized as follows. In Sec.II we briefly review the semiclassical approach to BOs in ordinary crystals, and highlight the difficulties to extend such a simple picture to BOs in lattices described by an effective non-Hermitian Hamiltonian. The exact analysis of BOs is presented in Sec.III, where an extension of the acceleration theorem is derived and the concept of complex trajectories is introduced. Two examples illustrating the onset of BOs in lattice models with a complex energy band are presented in Sec.IV, whereas the main conclusions are outlined in Sec.V.


\section{The semiclassical analysis of Bloch oscillations}
In this section we briefly review the simple semiclassical model that describes BOs as introduced for ordinary crystals, and show the limitations that arise when trying to extend such an analysis to complex crystals.  

\subsection{Single-band lattice model and effective Hamiltonian}
We consider the motion of a quantum particle in a one-dimensional periodic potential with lattice period $a$ driven by an external constant force $F$. As is well-known, for a weak force Zener tunneling among different bands is negligible and a single-band model can be used to describe the onset of BOs and the formation of WS localized states within each band \cite{r3bis,r18,r19}. Indicating by $W(x-na)$ the Wannier  state of the lattice band that localizes the particle at site $n$ of the lattice \cite{r3bis}, after expanding the particle wave function $\psi(x,t)$ as $\psi(x,t)=\sum_n c_n(t) W(x-na)$, the evolution equations for the site occupation probabilities read (taking $\hbar=1$)
\begin{equation}
i \frac{d c_n}{dt}=\sum_l \kappa_{l}c_{n+l}+nFac_n,
\end{equation}
where $F$ is the external driving force and $\kappa_{l}$  ($l \neq 0$) are the hopping rates among different lattice sites.The dispersion relation $E=E(q)$ of the lattice band is given by the relation
\begin{equation}
E(q)=\sum_{n} \kappa_{n} \exp(iqan)
\end{equation} 
where $- \pi/a \leq q < \pi/a$ is the Bloch wave number. Note that the constant $\kappa_0$ just provides an energy bias of the lattice band, and it will be taken to be zero in the following analysis for the sake of definiteness. In an Hermitian lattice, $\kappa_{-n}=\kappa_{n}^*$ and the energy dispersion curve $E(q)$ is a real function, whereas for a non-Hermitian lattice $E(q)$ can become complex. As is well known, the solution to the tight-binding equations (1) can be expressed as $c_n(t)=\phi(x=na,t)$, where the function $\phi(x,t)$ satisfies  the continuous Schr\"{o}dinger equation $i (\partial \phi / \partial t)=\hat{H}  \phi(x,t)$ with the {\it effective} Hamiltonian
\begin{equation}
\hat{H}=E \left( \hat{p}_x \right) +Fx
\end{equation}
where $\hat{p}_x= -i ( \partial / \partial x)$ is the momentum operator and $E(q)$ is the band dispersion relation defined by the Fourier expansion (2).
\subsection{Semiclassical analysis}
In an ordinary crystal, the onset of BOs for a wave packet with a narrow spectral distribution is generally explained on the basis of the Ehrenfest equations derived from  the effective Hamiltonian (3), that describe the evolution of mean position  (in units of the lattice period $a$) and momentum distributions of the particle wave function. Such equations read explicitly
\begin{eqnarray}
i \frac{d \langle x \rangle}{dt} & = & \langle [x, \hat{H}] \rangle= i \langle  \frac{\partial E}{\partial q} (\hat{p}_x) \rangle \\ 
i \frac{d \langle \hat{p}_x \rangle}{dt} & = & \langle [ \hat{p_x}, \hat{H}] \rangle=-i F.
\end{eqnarray}
From Eq.(5) it follows that the mean particle momentum $\langle \hat{p}_x(t) \rangle$ drifts in time with a constant speed $F$, i.e.
\begin{equation}
\langle \hat{p}_x(t) \rangle= \langle \hat{p}_x(0) \rangle-Ft
\end{equation}
a relation generally referred to as the 'acceleration theorem' in its simplest version \cite{r14}. Moreover, provided that the momentum distribution remains narrow in the evolution, the approximation $\langle  \frac{\partial E}{\partial q} (\hat{p}_x) \rangle \simeq  \frac{\partial E}{\partial q} ( \langle \hat{p}_x \rangle)$ in Eq.(4) can be introduced, yielding
\begin{equation}
\langle x(t) \rangle = \langle x(0) \rangle - \frac{1}{F} \left\{ E( \langle \hat{p}_x(t) \rangle)- E( \langle \hat{p}_x(0) \rangle) \right\}.
\end{equation}
 Since $E(q)$ is periodic with period $ 2 \pi/a$, i.e. $E(q+2 \pi/a)=E(q)$, it follows that in real space the particle wave packet undergoes an oscillatory motion, at the BO frequency $\omega_B=Fa$, along a path that reproduces the band dispersion curve of the lattice [according to Eq.(7)].\\
 The question is whether such a simple semiclassical analysis of BOs can be extended to the case of a complex lattice with a band dispersion relation $E(q)$ which is complex-valued, i.e. $E(q)=E_R(q)+iE_I(q)$ where  $E_R(q)$ and $E_I(q)$ are the real and imaginary parts, respectively, of the band dispersion curve. It should be noted that, as we will rigorously prove in the next section,  even though the energy spectrum $E(q)$ of the lattice is complex, as soon as $F \neq 0$ a real WS ladder energy spectrum is obtained, corresponding to a periodic motion at the BO frequency $\omega_B=Fa$ like in an ordinary crystal. However, the extension to the non-Hermitian case of the semiclassical analysis is a nontrivial issue and it is actually not much helpful. The reasons thereof are that, since the effective Hamiltonian $\hat{H}$ is now non-Hermitian, (i) the norm of the wave packet $ \langle \phi(t) | \phi(t) \rangle \equiv \int dx |\phi(x,t)|^2$ is not conserved, and (ii) the Ehrenfest equations are modified taking into account the commutation/anti-commutaion relations of the Hermitian and anti-Hermitian parts of $\hat{H}$. Namely, for any self-adjoint operator $\hat{A}$ corresponding to an observable $A$ that does not explicitly depend on time, after introduction of the expectation value of $\hat{A}$ defined as
\begin{equation}
\langle \hat{A}(t) \rangle \equiv \frac{\langle \phi | \hat{A} \phi \rangle}{\langle \phi | \phi \rangle}=\frac{\int dx \phi^*(x,t) \hat{A} \phi(x,t)}{\int dx |\phi(x,t)|^2}
\end{equation} 
 it can be readily shown that
 \begin{equation}
 i \frac{d}{dt} \langle \hat{A}(t) \rangle = \langle [\hat{A}, \hat{H}_1] \rangle+ \langle [\hat{A},\hat{H}_2]_+ \rangle -2 \langle \hat{A}(t) \rangle \langle \hat{H}_2 \rangle.
 \end{equation}
 To derive Eq.(9), we introduced the following decomposition of the effective Hamiltonian $\hat{H}$
 \begin{equation}
 \hat{H}=\hat{H}_1+\hat{H}_2
 \end{equation}
 as a sum of the Hermitian ($\hat{H}_1$) and anti-Hermitian ($\hat{H}_2$) parts, with \cite{note1}
 \begin{equation}
 \hat{H}_1=E_R(\hat{p}_x)+Fx \; ,\;\;\; \hat{H}_2=i E_I(\hat{p}_x)
 \end{equation}
 and indicated by $[\hat{A},\hat{B}]=\hat{A}\hat{B}-\hat{B}\hat{A}$ and by $[\hat{A},\hat{B}]_+=\hat{A}\hat{B}+\hat{B}\hat{A}$ the commutator and anti-commutator of operators $\hat{A}$ and $\hat{B}$, respectively.  The extension of the Eherenfest equations (4) and (5) to the non-Hermitian crystal is obtained from Eq.(9) by letting either $\hat{A}=x$ and $\hat{A}=\hat{p}_x$.  The resulting equations are rather cumbersome and are actually not much useful to predict the dynamical evolution of the particle wave packet. In particular, as compared to the Hermitian limit $\hat{H}_2=0$, additional terms, arising from the anticommutator and non-energy conservation terms in Eq.(9), do arise in the analysis, which completely destroy the simple solutions (6) and (7) and prevent to extend them in  the non-Hermitian case. As a matter of fact, in a non-Hermitian lattice with complex energy dispersion both the acceleration theorem [as least as stated by Eq.(6)] and Eq.(7) are violated, as discussed in the next section. 

\section{Bloch oscillations, acceleration theorem and complex trajectories}
A comprehensive analysis of BOs in non-Hermitian lattices with a complex energy dispersion curve should abandon the simple semiclassical model. The exact analysis, discussed in this section, shows that the following general results hold:\\
(i) For any applied force $F \neq 0$, the energy spectrum of the Hamiltonian is a WS ladder of equally-spaced levels (energy spacing $\omega_B=Fa$) like in an ordinary Hermitian crystal; thus the particle motion is periodic with a period 
\begin{equation}
T_B= \frac{2 \pi}{Fa};
\end{equation}
(ii) The acceleration theorem holds only {\it on average} in time, namely one has
\begin{equation}
\langle q(t) \rangle=\langle q(0) \rangle-Ft+\theta(t)
\end{equation}
  where $\langle q(t) \rangle$ is the mean value of the wave packet distribution in momentum space at time $t$ and $\theta(t)$ is a periodic function of 
  time with period $T_B$, $\theta(t+T_R)=\theta(t)$, and 
  $\theta(0)=0$; \\
  (iii) For a wave packet with initial narrow distribution in momentum space at around $q=0$, one has under certain smooth conditions
  \begin{equation}
  |\phi(x,t)|^2 \simeq G(t) |\phi(x-x_0(t),0)|^2
  \end{equation}
 where $x_0(t)$ is a path in the complex $x$ plane, defined by the relation
 \begin{equation}
 x_0(t)=\frac{E(0)-E(-Ft)}{F}
 \end{equation}
 whereas $G(t)$ is a periodic function with period $T_B$, describing norm oscillation within each BO cycle and given by
 \begin{equation}
 G(t) = \exp \left[ \frac{2}{F} \int_{-Ft}^{0} d \xi E_I( \xi) \right]. 
 \end{equation}

 \subsection{Wannier-Stark energy spectrum}
The exact analysis of BOs in non-Hermitian lattices can be done starting from the tight-binding equations (1) in the Wannier basis extending the procedure used for ordinary lattices. Let us introduced the spectrum $S(q,t)$ defined by the relation
\begin{equation}
S(q,t)=\sum_{n=-\infty}^{\infty} c_n(t) \exp(-i q n a).
\end{equation}
 which is a periodic function of $q$ with period $2 \pi /a$. Once the evolution of the spectrum $S(q,t)$ has been determined, the amplitude probabilities $c_n(t)$ are obtained after inversion of Eq.(17),namely one has
 \begin{equation}
 c_n(t)=\frac{a}{2 \pi} \int_{-\pi/a}^{\pi/a} dq S(q,t) \exp(iqna).
 \end{equation}
 From Eqs.(1), (2) and (17) it readily follows that the spectrum $S(q,t)$ satisfies the following differential equation
 \begin{equation}
i \frac{\partial S}{ \partial t}=E(q) S(q,t)+i F \frac{\partial S}{\partial q}.
 \end{equation}
 To determine the energy spectrum $\epsilon$ of the driven Hamiltonian, let us assume $c_n(t)=C_n \exp(-i \epsilon t)$, i.e. $S(q,t)=s(q) \exp(-i \epsilon t)$. Then from Eq.(19) it follows that 
 \begin{equation}
 i F \frac{\partial s}{\partial q}= \left[ \epsilon-E(q) \right]s
 \end{equation}
which can be solved for $s(q)$, yielding 
\begin{equation}
s(q)=s(0) \exp \left[  -i \frac{ \epsilon q}{F} +\frac{i}{F} \int_{0}^{q} d \xi E( \xi) \right].
\end{equation}
Since the spectrum $S(q,t)$ is a periodic function with respect to $q$ with period $2 \pi /a$, one has $s(2 \pi /a)=s(0)$, and thus from Eq.(21) it follows that the allowed energies $\epsilon$ satisfy the condition
\begin{equation}
-\frac{2 \pi \epsilon}{aF}+\frac{1}{F} \int_0^{2 \pi /a} d \xi E(\xi)=-2 l \pi
\end{equation}
where $l=0, \pm 1, \pm 2, \pm 3, ...$. Since $\int_0^{2 \pi /a} d \xi E(\xi)= 2 \pi \kappa_0 /a=0$ \cite{note2}, it follows that the energy spectrum is given by a Wannier-Stark ladder, i.e.
\begin{equation}
\epsilon_l=lFa \;\;\; (l=0, \pm 1, \pm 2, \pm 3, ...).
\end{equation}
The corresponding eigenstates, $C^{(l)}_n$, are readily obtained from Eqs.(18) and (21), and read explicitly
\begin{equation}
C^{(l)}_n= \mathcal{N}_l  \int_{-\pi/a}^{\pi/a} dq \exp \left[ iqa(n-l) +\frac{i}{F} \int_0^q d \xi E( \xi)   \right]
\end{equation}
 where $\mathcal{N}_l$ is a normalization constant.
 
 \subsection{Acceleration theorem}
 The extension of the 'acceleration theorem' to non-Hermitian lattices is obtained by looking at the general solution to the spectral equation (19), which reads explicitly
 \begin{equation}
 S(q,t)=S_0(q+Ft) \exp \left[ -i \int_0^t d \xi E(q+Ft-F \xi) \right]
 \end{equation}
 where $S_0(q)$ is the spectrum of the particle wave packet at initial time $t=0$, i.e. $S_0(q)=S(q,0)$. Clearly, in an Hermitian lattice in which $E(q)$ is a real function from Eq.(25) one has $|S(q,t)|^2=|S_0(q+Ft)|^2$, i.e. the particle wave packet in the momentum space drifts undistorted with a constant speed equal to the force $F$ (acceleration theorem). In a non-Hermitian lattice with complex energy dispersion curve $E(q)$ such a result {\it does not hold}, since one has in this case
 \begin{equation}
 |S(q,t)|^2=|S_0(q+Ft)|^2 \exp \left[ 2 \int_0^t d \xi E_I(q+Ft-F \xi) \right]
 \end{equation}
It is worth determining the motion of the center of mass $\langle q(t) \rangle$ of the spectrum, defined by the relation
\begin{equation}
\langle q(t) \rangle = \frac{\int_{-\pi/a}^{\pi/a} dq q |S(q,t)|^2}{\int_{-\pi /a}^{\pi /a} dq |S(q,t)|^2}.
\end{equation}
From Eqs.(26) and (27) it readily follows that \cite{note3}
\begin{equation}
\langle q(t) \rangle=\langle q(0) \rangle-Ft+ \theta(t)
\end{equation}
where $\theta(t)$ is a periodic function of period $T_B$, i.e. $\theta(t+T_B)=\theta(t)$, with $\theta(0)=0$, given by
\begin{eqnarray}
\theta(t) & = &  \frac{ \int_{-\pi/a}^{\pi/a} dq q |S_0(q)|^2  \exp \left[  2 \int_0^t d \xi E_I(q-F \xi) \right]}{\int_{-\pi/a}^{\pi/a} dq |S_0(q)|^2   \exp \left[  2 \int_0^t d \xi E_I(q-F \xi) \right] } \nonumber \\
& - & \frac{ \int_{-\pi/a}^{\pi/a} dq q |S_0(q)|^2}{\int_{-\pi/a}^{\pi/a} dq |S_0(q)|^2} 
\end{eqnarray}
Equation (28) shows that, in addition to the constant drift of the particle momentum $ \langle q(t) \rangle$ found in the Hermitian limit, an oscillatory motion is superimposed, i.e. the drift occurs only on average over each BO cycle. 
 
 \subsection{Bloch oscillations and complex trajectories}
 The evolved amplitude probabilities $c_n(t)$ at time $t>0$ are determined from the initial spectral distribution $S_0(q)$ of the particle wave packet and can be determined from Eqs.(17), (18) and (25). One obtains 
 \[
 c_n(t)=\sum_l U_{n,l}(t) c_l(0)
 \] 
where the propagator is given by
\begin{eqnarray}
U_{n,l}(t) =  \frac{a}{2 \pi} \exp(-iFant) \times  \;\;\;\;\;\;\;\;\;\;\;\; \\
   \int_{-\pi/a}^{\pi/a} dq \exp \left[  -i \int_0^t d \xi E(q-F \xi) +i q (n-l)a \right] \;\; \nonumber
\end{eqnarray}
Let us now consider a particle wave packet with an initial narrow distribution in momentum space, i.e. let us assume that $|S_0(q)|^2=|S(q,0)|^2$ is a narrow function of $q$ peaked at around $q=q_0$, falling off sufficient rapidly far from $q=q_0$. For the sake of simplicity, in the following we will consider the case $q_0=0$. In this case, the energy dispersion relation $E(q)$ in the exponent on the integral on the right hand side of Eq.(30) can be expanded in series of $q$ at around $q=0$,  
 \begin{equation}
 E(q-F \xi) \simeq E(-F \xi)+q \left( \frac{d E}{ d q} \right)_{(-F \xi)}+ ... \;\;,
 \end{equation}
  and the integral can be extended from $q= -\infty$ to $q= \infty$. If we limit the expansion in Eq.(31) up to first order in $q$, after setting $c_n(t)=\phi(x=na,t)$ one readily obtains \cite{note4}
  \begin{equation}
  \phi(x,t) \simeq  \exp \left[ -iFx -i \int_0^t d \xi E(-F \xi) \right] \phi(x-x_0(t),0)
  \end{equation}
  where the complex path $x_0(t)$ is defined by Eq.(15). From Eq.(32) it follows that
  \begin{equation}
  |\phi(x,t)|^2  \simeq  G(t) |\phi(x-x_0(t),0)|^2
  \end{equation}
  where $G(t)$ is defined by Eq.(16) \cite{note5}. Hence, apart from an oscillation of the amplitude over each BO cycle as defined by the periodic function $G(t)$, in real space the particle wave packet undergoes an oscillatory motion described by the orbit $x_0(t)$ in {\it complex} plane. Note that the trajectory of the wave packet center of mass in real space is not simply related to $x_0(t)$, and it generally depends on the specific profile $\phi(x,0)$. For some initial profiles, e.g. for a Gaussian wave packet distribution $\phi(x,0) = \mathcal {N} \exp(-\gamma x^2)$ with $\gamma$ real, from Eq.(33) it turns out that the trajectory in real space is simply given by the real part of $x_0(t)$, i.e. it is determined by the real part of the energy dispersion curve (like in the Hermitian lattices). However, this is not a general rule, as discussed in the next section for a specific non-Hermitian lattice model.
  
\begin{figure}[htbp]
  \includegraphics[width=84mm]{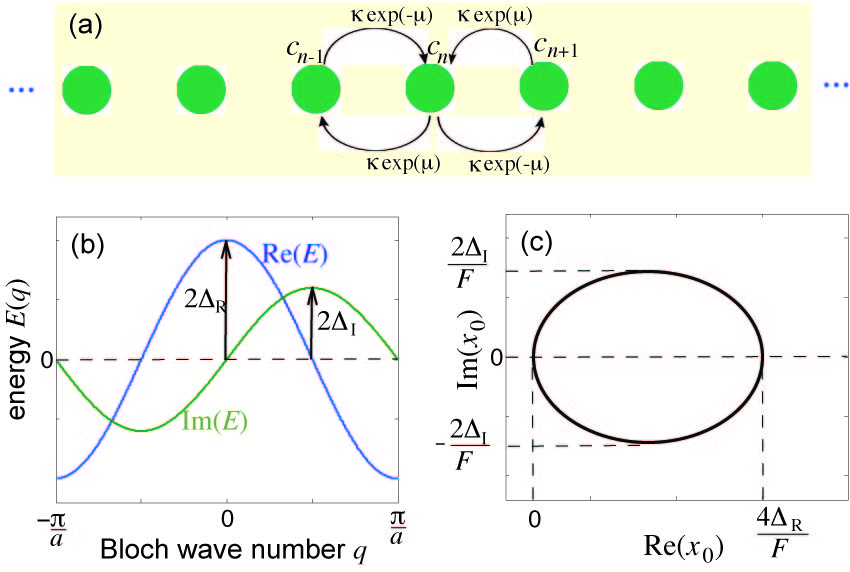}\\
   \caption{(color online) (a) Schematic of a one-dimensional tight-binding chain with an imaginary gauge field (Hatano-Nelson model), and (b) corresponding band diagram (real and imaginary parts of the band dispersion curve). (c) Closed orbit in complex $x$ plane (ellipse) followed by a particle wave packet  under a forcing $F$.}
\end{figure}

\begin{figure}[htbp]
  \includegraphics[width=84mm]{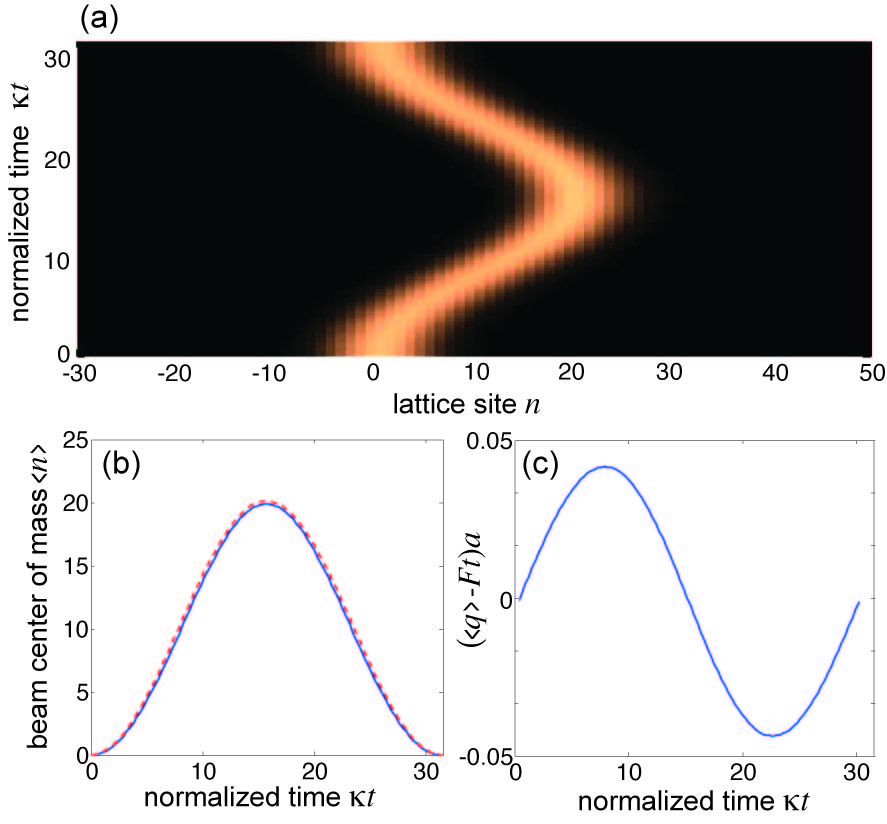}\\
   \caption{(color online) (a) Propagation of a Gaussian wave packet (snapshots of $|c_n(t)|^2 / \sum_n |c_n(t)|^2$) in the Hatano-Nelson lattice over one BO cycle for parameter values $Fa/ \kappa=0.2$ and $\mu=0.1$. In (b) and (c) the trajectories of the beam center of mass $\langle n(t) \rangle$ and   $\langle q(t) \rangle$ in real and momentum space, over one BO cycle, are shown by solid lines. The dashed curve in (b), almost overlapped with the solid one, is the beam trajectory in real space as predicted by Eqs.(14) and (15), which turns out to be given simply by ${\rm Re}(x_0)$, where $x_0(t)$ is the elliptical orbit in complex plane depicted in Fig.1(c). Breakdown of the acceleration theorem is clearly shown in panel (c), where the mean momentum $\langle q(t) \rangle$ is not a linear function of time owing to a small correction $\theta(t)$ according to Eqs.(28) and (29). }
\end{figure}

\begin{figure}[htbp]
  \includegraphics[width=84mm]{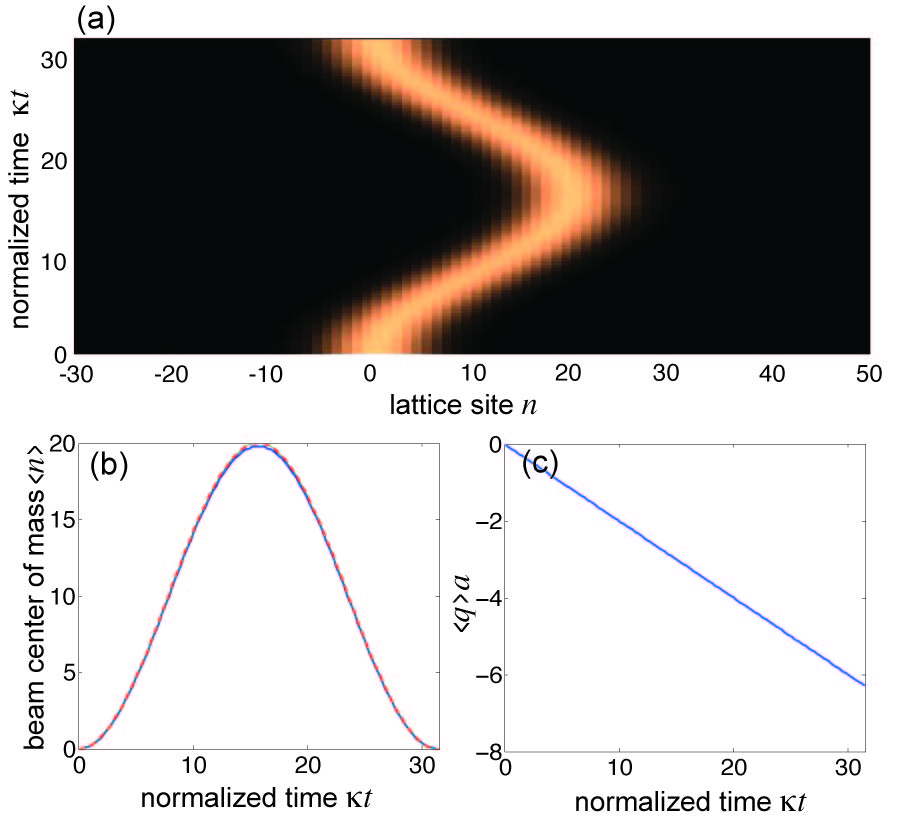}\\
   \caption{(color online) Same as Fig.3, but in the Hermitian limit ($\mu=0$).}
\end{figure}

  \section{Examples of Bloch oscillations with complex trajectories}
  To exemplify the concept of BOs with complex trajectories, we consider two exactly-solvable examples of non-Hermitian lattices with a complex band dispersion relation.\par
  The first example is provided by the Hatano-Nelson non-Hermitian tight-binding lattice, which describes the hopping motion of a quantum particle on a one-dimensional tight-binding lattice in the presence of an imaginary vector field; see Fig.1(a).  In the nearest-neighbor approximation, it is described by the following set of equations for the site occupation probabilities $c_n(t)$
  \begin{equation}
  i \frac{dc_n}{dt}= \kappa \exp(\mu) c_{n+1}+ \kappa \exp(-\mu) c_{n-1}+V_n c_n
  \end{equation}
  where $\kappa>0$ is the hopping rate between adjacent lattice sites, $\mu>0$ is the imaginary vector potential, and $V_n$ are the on-site energies. The model (34) was originally introduced by Hatano and Nelson in a pioneering work to study the motion of magnetic 
flux lines in disordered type-II superconductors \cite{r33_1}, showing that an 'imaginary' gauge field in a disordered one-dimensional lattice can induce a delocalization transition, i.e. it can prevent Anderson localization \cite{r33_1,r33_2,r33_3}. A possible implementation in an optical setting of the Hatano-Nelson model (34), based on light transport in coupled resonator optical waveguides, has been recently suggested in Ref.\cite{r33_4}. Here we consider the case of an ordered lattice with $V_n=Fan$, so that the lattice model (34) is obtained from the general model described by Eq.(1) after setting 
  \begin{equation}
  \kappa_l= \kappa \exp( \mu) \delta_{l,1}+\kappa \exp(-\mu) \delta_{l,-1}.
  \end{equation}
  Note that non-Hermiticity in the Hatano Nelson model arises from the imaginary gauge field ($ \mu \neq 0$), the Hermitian case being reached in the limit $\mu=0$ \cite{note6}. The energy dispersion relation of the lattice band reads [Fig.1(b)]
  \begin{equation}
  E(q)=2 \Delta_R \cos(qa)+2 i \Delta_I \sin(qa)
  \end{equation}
  where we have set
  \begin{equation}
  \Delta_R= \kappa \cosh( \mu) \; ,\;\;\; \Delta_I= \kappa \sinh( \mu).
  \end{equation}
   An interesting feature of the Hatano-Nelson model is that it can be mapped into an Hermitian lattice problem after the substitution $c_n(t)=a_n(t) \exp( -\mu n)$ in Eq.(34). Such a property can be readily used to derive analytical expression of the WS eigenstates $C_n^{(l)}$, which are given by [according to Eq.(24)]
   \begin{equation}
   C_n^{(l)}= \mathcal{N}_l \exp[- \mu n+i \pi (n-l)] J_{n-l} \left( \frac{2 \kappa}{Fa} \right),
   \end{equation} 
   whereas the propagator $U_{n,l}(t)$, as obtained from Eq.(30), reads explicitly
   \begin{eqnarray}
   U_{n,l}(t) & = & J_{n-l} \left(  \frac{4 \kappa}{Fa} \sin \left( \frac{Fat}{2} \right)  \right) \exp(-iFnat) \times \nonumber \\
   & \times & \exp \left[ i(n-l) (Fat-\pi)/2+ 	\mu(l-n) \right]
   \end{eqnarray}
   where $J_n$ is the Bessel function of first kind and order $n$.
   According to Eq.(15), a particle wave packet with narrow distribution in momentum space undergoes an oscillatory motion with a trajectory in complex plane described by the equation
 \begin{equation}
 x_0(t)=\frac{2 \Delta_R}{F} \left[ 1-\cos(Fat) \right]+i \frac{2 \Delta_I}{F} \sin(Fat)
 \end{equation}  
   which describes the orbit of an ellipse [see Fig.1(c)]. As an example, Fig.2 shows the numerically-computed propagation of an initial Gaussian wave packet with distribution $c_n(0) \propto  \exp(-\gamma n^2)$ for parameter values $Fa/\kappa=0.2$, $\gamma=0.02$ and $\mu=0.1$. In the figure, the evolution of the wave packet center of mass $\langle n (t) \rangle$ and $\langle q(t) \rangle$ in real and momentum space, over one BO cycle, are also depicted. For comparison, the limit of an Hermitian lattice ($\mu=0$) is shown in Fig.3. Note that in the non-Hermitian case the mean momentum $\langle q(t) \rangle$ does not exactly varies linearly in time because of a non vanishing (thought small) variation $\theta(t)$, as shown in Fig.2(c). Note also that for a Gaussian wave packet, according to Eqs.(14) and (15) the behavior of $\langle n(t) \rangle$ is simply given by
   \begin{equation}
   \langle n(t) \rangle={\rm Re} \left( x_0(t) /a \right) 
   \end{equation}
    i.e. the trajectory of the wave packet in the real space over one BO cycle basically maps the {\it real part} of the band dispersion curve $E_R(q)$. At first sight one might think that this should be a general result, since the real part of the band dispersion curve is associated to the group velocity of the wave packet. However, this is not the case since the semiclassical model can not be readily extended to the non-Hermitian case, as discussed in Sec.II. For example, let us consider a two-humped complex initial distribution defined by the relation
    \begin{equation}
    c_n(0) 	\propto \frac{1}{\cosh^2[(\alpha+i \beta) n]}
    \end{equation}    
\begin{figure}[htbp]
  \includegraphics[width=84mm]{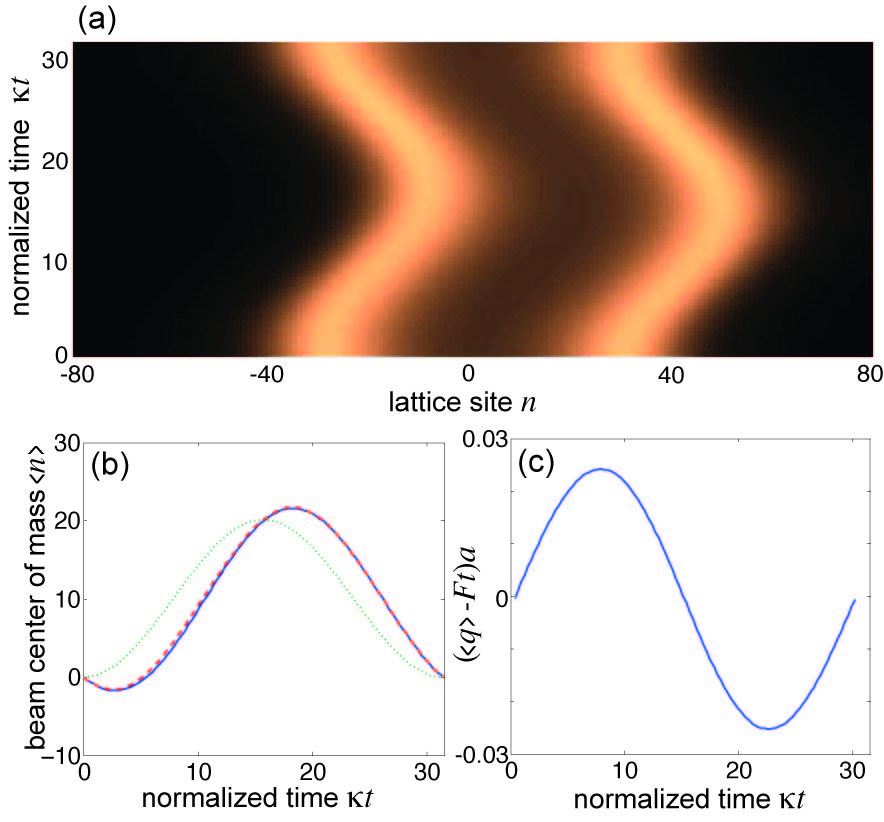}\\
   \caption{(color online) Same as Fig.2, but for an initial two-humped wave packet distribution defined by Eq.(42). Parameter values are given in the text. In panel (b) the dashed curve, almost overlapped with the solid one, corresponds to the trajectory of the wave packet center of mass as predicted by Eqs.(14) and (15), whereas the tiny dotted curve shows the behavior of ${\rm Re}(x_0(t)/a)$.}
\end{figure}
\begin{figure}[htbp]
  \includegraphics[width=84mm]{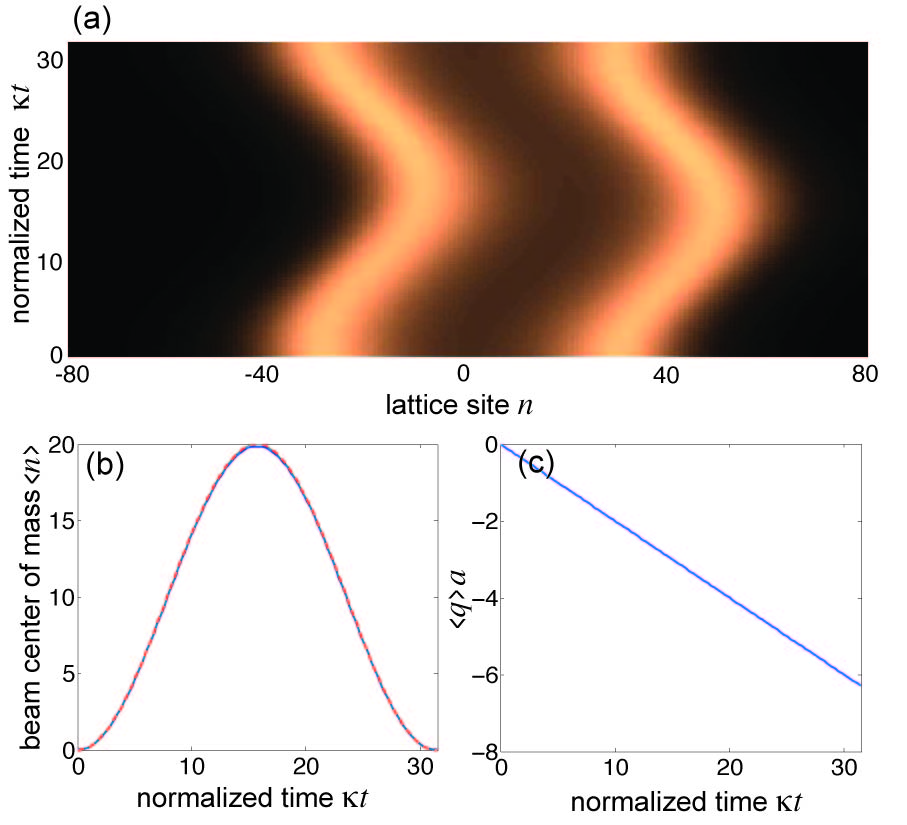}\\
   \caption{(color online) Same as Fig.4, but in the Hermitian limit ($\mu=0$).}
\end{figure}
\begin{figure}[htbp]
  \includegraphics[width=84mm]{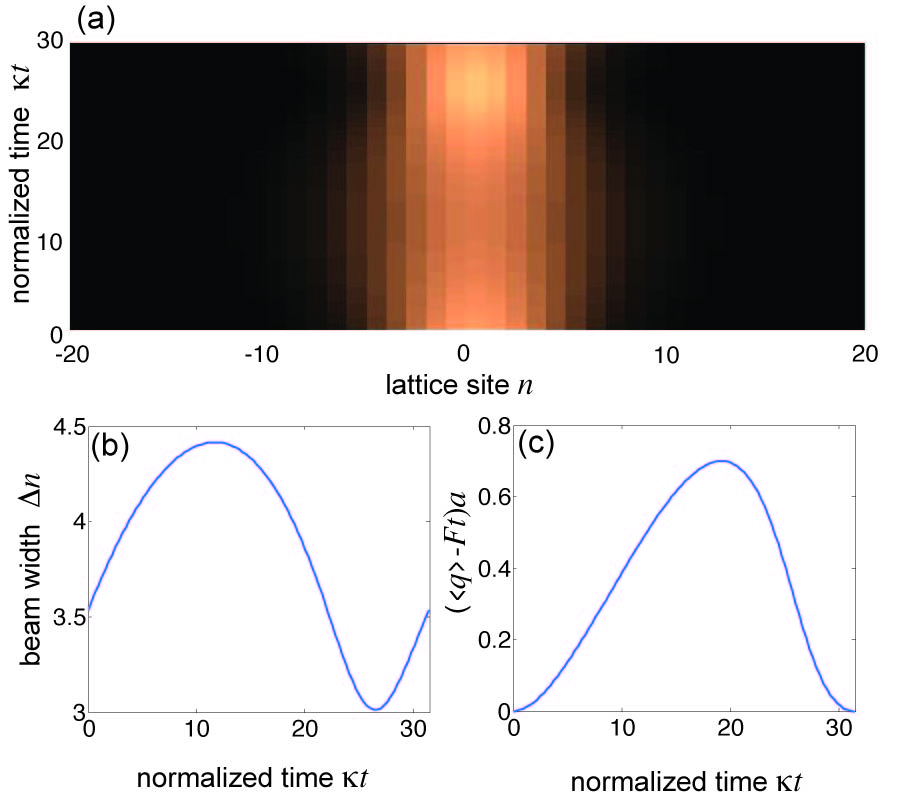}\\
   \caption{(color online) (a) Propagation of a Gaussian wave packet (snapshots of $|c_n(t)|^2 / \sum_n |c_n(t)|^2$) over one BO cycle in the non-Hermitian lattice defined by Eqs.(1) and (44)  for $Fa/ \kappa=0.2$. The initial condition is $c_n(0) \propto \exp(-\gamma n^2)$ with $\gamma=0.02$. In (b)  the behavior of the wave packet width $\Delta n(t)$, defined as $\Delta n(t)= \sqrt{\sum_n n^2 |c_n(t)|^2 / \sum_n  |c_{n}(t)|^2}$, is shown, indicating a breathing (rather than oscillatory) dynamics of the wave packet. The behavior of the centroid in momentum space,  $\langle q(t) \rangle$, is shown in panel (c).}
\end{figure}
     with $\alpha$ and $\beta$ real parameters. A narrow wave packet distribution at around $q=0$ in momentum space is ensured by assuming $|\alpha|, |\beta| \ll 1$.  Figure 4 shows the numerically-computed wave packet evolution for the initial condition (42) and for parameter values $Fa / \kappa=0.2$, $\alpha=0.02$, $\beta=0.04$, and $\mu=0.1$. The Hermitian limit $\mu=0$ is shown, for comparison, in Fig.5. While in the Hermitian limit the wave packet trajectory in the real space maps the energy band dispersion curve, in the non-Hermitian case the trajectory does not merely correspond to Eq.(41).\\
     As a second example, let us consider a non-Hermitian lattice with a completely imaginary dispersion relation, given by
     \begin{equation}
     E(q)=2 i \kappa \cos(qa)
     \end{equation}
     with $\kappa>0$. Such a lattice model  is obtained from the general model described by Eq.(1) after setting 
  \begin{equation}
  \kappa_l= i \kappa  \delta_{l,1}+i \kappa \delta_{l,-1}
  \end{equation}
  i.e. by taking imaginary hopping rates for nearest neighbor sites. Such a model  can be realized, for example, in optical waveguide lattices with synthetic complex hopping rates obtained by longitudinal gain/loss management \cite{r33_8,r41} or in optical fiber loops with an amplitude modulator \cite{r41,r33_9}.   
  In this case, the orbit $x_0(t)$ in the complex plane is simply given by
  \begin{equation}
  x_0(t)=(2 \kappa /F) i [1-\cos(Fat)],
  \end{equation}
  i.e. it is a straight line along the imaginary axis. The propagator $U_{n,l}(t)$, as obtained from Eq.(30), reads explicitly
   \begin{eqnarray}
   U_{n,l}(t) & = & I_{n-l} \left(  \frac{4 \kappa}{Fa} \sin \left( \frac{Fat}{2} \right)  \right) \exp(-iFnat) \times \nonumber \\
   & \times & \exp \left[ i(n-l) Fat /2 \right].
   \end{eqnarray}
where $I_n$ is the modified Bessel function of order $n$.

   Interestingly, for an initial Gaussian distribution $c_n(0) \propto \exp(-\gamma n^2)$ the center of mass of the wave packet in real space does not undergo any oscillation, i.e. $\langle n(t) \rangle=0$;  the purely imaginary trajectory $x_0(t)$ in complex plane corresponds this time to a breathing dynamics of the wave packet, as shown in Fig.6. Hence rather generally the fact that in non-Hermitian lattices with a complex dispersion curve the trajectory of a wave packet occurs in complex plane means that a combined oscillatory and breathing dynamics can occur, and that such effects may depend on the specific initial wave packet profile. 
   
   \section{Conclusion} 
   In this work we have theoretically investigated the onset of BOs in non-Hermitian lattices with a complex energy band, highlighting a few distinctive and novel features as compared to BOs in lattices (either Hermitian or non-Hermitian) with an entirely real energy band. Specifically, we have shown that the simple semiclassical model of BOs, based on the Ehrenfest equations of motion for an effective Hermitian Hamiltonian, can not be extended to non-Hermitian lattices with a complex energy band. In particular, we have shown that the so-called 'acceleration theorem' must be extended to account for reshaping of the  wave packet  distribution in momentum space, and that the particle wave packet in real space undergoes a periodic motion describing rather generally a complex trajectory, i.e. it generally corresponds to reshaping and breathing of the wave packet in addition to a transverse oscillatory motion. The concept of BOs with complex trajectories has been exemplified by considering two examples of non-Hermitian lattices with a complex band dispersion relation, namely the Hatano-Nelson tight-binding Hamiltonian describing the hopping motion of a quantum particle on a linear lattice with an imaginary vector potential and the nearest-neighbor tight-binding lattice Hamiltonian with imaginary hopping rates. Our results shed new light onto the coherent transport properties in driven non-Hermitian crystals, and the predictions of the analysis could be observed in synthetic temporal or spatial crystals realized with optical structures \cite{r39,r33_4bis,r33_9}. Here we have focused our study to one-dimensional non-Hermitian lattices driven by a dc force, however it would be interesting to extend the analysis to ac and dc-ac driven lattices, to bi-dimensional lattices, lattices with synthetic gauge fields, etc. (see, for instance, \cite{r42,r43,r44} and references therein). In such systems, it will be interesting to investigate the impact of a non-Hermitian effective lattice band to such important phenomena like dynamic localization, trapping and unidirectional transport.


\begin{thebibliography}{}
\bibitem{note2}
For a non vanishing value of $\kappa_0$, the imaginary part of $\kappa_0$ is responsible for either a damping or amplification of BOs and the WS energy spectrum acquires a non vanishing imaginary part, equal to all modes
\bibitem{note3}
 The right hand side in Eq.(28) has to be computed mod. $ 2 \pi /a$ inside the interval $(- \pi/a, \pi/a)$
\bibitem{note4}
 The series expansion approximation holds provided that $|S_0(q)|$ vanishes far from $q=0$ faster than exponential, namely provided that $|S_0(q)| \exp(M|q|) \rightarrow 0$ as $|q| \rightarrow \infty$, where $M= {\max}_{(-\pi/a,\pi/a)} E_I(q)$
\bibitem{note5}
 In case of an initial wave packet with spectral distribution $S_0(q)$ is momentum space narrow at around $q=q_0 \neq 0$, Eqs.(15) and (16) are generalized as follows 
 \[
 x_0(t) = \frac{E(q_0)-E(q_0-Ft)}{F}
 \]
 \[ G(t)=\exp \left[ \frac{2}{F} \int_{-Ft}^{0} d \xi E(q_0+ \xi) \right]
 \]
\end{thebibliography}
\end{document}